\documentclass[a4paper,11pt,final]{scrartcl}
\usepackage[utf8]{inputenc}
\usepackage{ifpdf}
\usepackage{latexsym}
\usepackage{exscale}
\usepackage{amsfonts}
\usepackage{eucal}
\usepackage{cmmib57}
\usepackage[centertags]{amsmath}
\usepackage{amssymb}
\usepackage{amsbsy}
\usepackage{amstext}
\usepackage{amsthm,thmtools}
\usepackage{xcolor}
\usepackage{framed}
\usepackage{tocbibind}
\usepackage{tocloft}
\usepackage{etoolbox}
\usepackage[ruled,vlined,algosection]{algorithm2e}
\usepackage{delarray}
\usepackage[automark]{scrpage2}
\usepackage{times}
\usepackage[toc,page]{appendix}
\usepackage{textcomp}
\usepackage{mflogo}
\usepackage{here}
\usepackage[round]{natbib}
\usepackage{here}
\usepackage{longtable}
\usepackage{threeparttablex}
\usepackage{multirow}
\usepackage{multicol}
\usepackage{caption}
\usepackage{lscape}
\usepackage{rotating,capt-of}
\usepackage[margin=15mm]{geometry}
\usepackage{tikz}
\usetikzlibrary{shapes}

\newcounter{treeline}

\ifvtexpdf\pdftrue\fi
\ifpdf
\usepackage{graphicx}
\usepackage[pdftex]{thumbpdf}
\usepackage{nameref}
\usepackage[pdftex,colorlinks]{hyperref}
\hypersetup{
           breaklinks=true,
           pdfpagemode=UseOutlines,
           colorlinks=true,
           linkcolor=blue,
           citecolor=blue,
           bookmarksopen=true,
           pdftitle={The Incorrect Usage of Propositional Logic in Game Theory},
           pdfauthor={Holger Ingmar Meinhardt},
           pdfsubject={Tu Games, Propositional Logic},
           pdfkeywords={Axiomatization, Propositional Logic, Material Implication, Indirect Proof, Tu Games},
}

\else
\usepackage{epsfig}
\usepackage[ps2pdf]{thumbpdf}
\usepackage{nameref}
\usepackage[ps2pdf]{hyperref}

\hypersetup{
           breaklinks=true,
           pdfpagemode=UseOutlines,
           colorlinks=true,
           linkcolor=blue,
           citecolor=blue,
           bookmarksopen=true,
           pdftitle={The Incorrect Usage of Propositional Logic in Game Theory},
           pdfauthor={Holger Ingmar Meinhardt},
           pdfsubject={Matlab Game Theory Toolbox, Tu Games},
           pdfkeywords={Axiomatization, Propositional Logic, Material Implication, Indirect Proof, Tu Games},
}
\fi

\addtocounter{MaxMatrixCols}{20}

\makeatletter
\renewcommand{\section}{\@startsection
  {section}%
  {1}%
  {0em}%
  {-\baselineskip}%
  {0.5\baselineskip}%
  {\centering\normalfont\Large\scshape\mdseries}}%
\makeatother

\makeatletter
\renewcommand{\subsection}{\@startsection
  {subsection}%
  {2}%
  {0em}%
  {-\baselineskip}%
  {0.5\baselineskip}%
  {\normalfont\large\scshape\mdseries}}%
\makeatother

\makeatletter
\renewcommand*\env@matrix[1][c]{\hskip -\arraycolsep
  \let\@ifnextchar\new@ifnextchar
  \array{*\c@MaxMatrixCols #1}}
\makeatother

\makeatletter

\newenvironment{theopargself*}
    {\def\@spopargbegintheorem##1##2##3##4##5{\trivlist
         \item[\hskip\labelsep{##4##1\ ##2}]{\hspace*{-\labelsep}##4##3\@thmcounterend}##5}
     \def\@Opargbegintheorem##1##2##3##4{##4\trivlist
         \item[\hskip\labelsep{##3##1}]{\hspace*{-\labelsep}##3##2\@thmcounterend}}}{}
\def \@floatboxreset {%
        \reset@font
        \small
        \@setnobreak
        \@setminipage
}
\def\figure{\@float{figure}}
\@namedef{figure*}{\@dblfloat{figure}}
\def\table{\@float{table}}
\@namedef{table*}{\@dblfloat{table}}
\def\fps@figure{htbp}
\def\fps@table{htbp}

\makeatother

\makeatletter
\@addtoreset{equation}{section}
\makeatother
\makeatletter
\@addtoreset{table}{section}
\makeatother

\theoremstyle{plain}

\newtheoremstyle{break}% name
  {9pt}%      Space above, empty = `usual value'
  {9pt}%      Space below
  {\itshape}% Body font
  {}%         Indent amount (empty = no indent, \parindent = para indent)
  {\bfseries}% Thm head font
  {.}%        Punctuation after thm head
  {\newline}% Space after thm head: \newline = linebreak
  {}%         Thm head spec

\newtheoremstyle{break1}% name
  {9pt}%      Space above, empty = `usual value'
  {9pt}%      Space below
  {\rmfamily}% Body font
  {}%         Indent amount (empty = no indent, \parindent = para indent)
  {\scshape}% Thm head font
  {.}%        Punctuation after thm head
  {\newline}% Space after thm head: \newline = linebreak
  {}%         Thm head spec

\theoremstyle{break}

\newtheoremstyle{note}% name
  {3pt}%      Space above
  {3pt}%      Space below
  {}%         Body font
  {}%         Indent amount (empty = no indent, \parindent = para indent)
  {\itshape}% Thm head font
  {:}%        Punctuation after thm head
  {.5em}%     Space after thm head: " " = normal interword space;
  {\newline}  %       \newline = linebreak
  {}%         Thm head spec (can be left empty, meaning `normal')

\theoremstyle{note}

\theoremstyle{definition}
\newtheorem{example}{Example}[section]

\theoremstyle{break1}

\begin{document}
\bibliographystyle{plainnat} %,abbrvnat,unsrtnat
\pdfbookmark[0]{The Incorrect Usage of Propositional Logic in Game Theory: The Case of Disproving Oneself}{tit}
\title{{The Incorrect Usage of Propositional Logic in Game Theory: The Case of Disproving Oneself} 
}
\author{{\bfseries Holger I. MEINHARDT} %~\thanks{We are very grateful to Martha {Saboy\a'a} Baquero and to two anonymous referees for their comments and suggestions of improvements on how to improve the presentation of the paper. Of course, the usual disclaimer applies.}
~\thanks{Holger I. Meinhardt, Institute of Operations Research, Karlsruhe Institute of Technology (KIT), Englerstr. 11, Building: 11.40, D-76128 Karlsruhe. E-mail: \href{mailto:Holger.Meinhardt@wiwi.uni-karlsruhe.de}{Holger.Meinhardt@wiwi.uni-karlsruhe.de}} 
}
%\date{August 29, 2011}
\maketitle

\begin{abstract}
Recently, we had to realize that more and more game theoretical articles have been published in peer-reviewed journals with severe logical deficiencies. In particular, we observed that the indirect proof was not applied correctly. These authors confuse between statements of propositional logic. They apply an indirect proof while assuming a prerequisite in order to get a contradiction. For instance, to find out that ``{\itshape if $A$ then $B$}'' is valid, they suppose that the assumptions ``$A$ and not $B$'' are valid to derive a contradiction in order to deduce ``{\itshape if $A$ then $B$}''. Hence, they want to establish the equivalent proposition ``{\itshape $A \land$ not $B$} implies $A \land not A$'' to conclude that ``{\itshape if $A$ then $B$}''is valid. In fact, they prove that a truth implies a falsehood, which is a wrong statement. As a consequence, ``{\itshape if $A$ then $B$}'' is invalid, disproving their own results. We present and discuss some selected cases from the literature with severe logical flaws, invalidating the articles.\\

\noindent {\bfseries Keywords}: Transferable Utility Game, Solution Concepts, Axiomatization, Propositional Logic, Material Implication, Circular Reasoning (circulus in probando), Indirect Proof, Proof by Contradiction, Proof by Contraposition, Cooperative Oligopoly Games \\%[.25em]

\noindent {\bfseries 2010 Mathematics Subject Classifications}: 03B05, 91A12, 91B24  \\
\noindent {\bfseries JEL Classifications}: C71 % D43, L13
\end{abstract}

%%%%%%%%%

\thispagestyle{empty}
\pagebreak

\pagestyle{scrheadings}  \ihead{\empty} \chead{The Incorrect Usage of Propositional Logic in Game Theory} \ohead{\empty}

\section{Introduction}
During the last decades, game theory has encountered a great success while becoming the major analysis tool for studying conflicts and cooperation among rational decision makers. We observed fine and groundbreaking works based on solid and rigorous mathematical propositions and thinking. However, recently, we discovered that more and more articles have been published in peer-reviewed journals with severe fallacies. Especially, we had to learn that the indirect proof, which is based on a material implication, was not applied correctly. A material implication is a rule of replacement that allows to replace a conditional statement by a disjunction. These authors confuse and mix up non-equivalent fundamental statements from propositional logic to come up with a desired contradiction without asking if the derived conclusion makes sense from a logical point of view. 

A statement {\itshape if $A \Rightarrow B$} and its contrapositive {\itshape if $\neg B \Rightarrow \neg A$} are logically equivalent statements, which are also equivalent to the disjunction $\neg A \lor B$. The common proof technique based on a material implication replaces the conditional statement {\itshape if $A \Rightarrow B$} by the disjunction $\neg A \lor B$. It should be evident that the conjunction $A \land \neg B$ is the negation of the disjunction $\neg A \lor B$, and that it is not its contrapositive. For instance, to prove the implication {\itshape if $A \Rightarrow B$}, we can focus on the opposite {\itshape $\neg (A \Rightarrow B)\equiv \neg (\neg A \lor B) \equiv$ $(A \land \neg B)$} in order to get from {\itshape if $A \Rightarrow B$} the logical equivalent implication {\itshape if $A \land \neg B \Rightarrow B \land \neg B$}. This imposes a proof by contradiction, since $B \land \neg B$ is a falsum $\bot$. However, if the starting point is a proof by contraposition, i.e., $\neg B \Rightarrow \neg A$, we obtain the following equivalent statement $A \land \neg B \Rightarrow A \land \neg A$. It should be evident that this also imposes a proof by contradiction.    

In accordance with $(A \land \neg B \Rightarrow A \land \neg A) \equiv (A \Rightarrow B)$, one has to be careful concerning the logical conclusions when combining a proof by contradiction with a material implication. To get a valid proposition, one has to assume that $A \land \neg B$ is an invalid premise ($\neg A \lor B$ valid) from which a false statement like $A \land \neg A$ can be deduced. Then, we know that the implication $A \land \neg B \Rightarrow A \land \neg A$ is a valid statement, and from this result, we can infer that the original statement $A \Rightarrow B$ is also a truth. However, a wrong proposition is obtained while assuming first $A \land \neg B$ to be a true premise ($\neg A \lor B$ invalid), and then deriving the falsehood $A \land \neg A$. Here, one derives from a true premise something what is false. This statement is obviously a falsehood. As a consequence, one can infer that $A \Rightarrow B$ is invalid, i.e., $A \not\Rightarrow B$.        

In contrast, it is not a permissible conduct to derive from a valid premise $A \land \neg B$ a so-called contradiction, say $\neg A$, to deduce that $A \land \neg B$ is false, and from this outcome, one follows that the negation of $A \land \neg B$, i.e., the disjunction $\neg A \lor B$ must be valid, and therefore $A \Rightarrow B$ must follow too. This is a fallacy. Actually, one has established that something true implies something which is false. This is an incorrect implication. Doing so, disproves the result. 

Similar, it is a fallacy to assume that $A \Rightarrow B$ is false, i.e., $A \land \neg B$ holds in order to derive a contradiction, say $A \land \neg A$, to finally deduce from this contradiction that $A \land \neg B$ is false, and that one has therefore proved $A \Rightarrow B$ by the logical equivalence of $A \land \neg B \Rightarrow A \land \neg A$ and $A \Rightarrow B$. Again, one has disproved oneself, since one gets that $A \land \neg B \Rightarrow A \land \neg A$ is a falsehood confirming that $A \Rightarrow B$ is false as well. Obviously, this kind of arguing is a circular reasoning (circulus in probando). Unfortunately, this is exactly the line of argument that we have observed in our sample from the literature. These authors have shown in their proofs the exact opposite of what had been intended to prove. 

To summarize, the authors try to establish that a proposition $\phi$ ``satisfies'' a falsum $\bot $ to conclude that $\neg \phi$ holds, i.e., $(\phi \vdash \bot) \Leftrightarrow \neg \phi$. This constitutes a formal expression of an indirect proof. However, it should be evident that this is not the same as $(\phi \Rightarrow \bot) \Leftrightarrow \neg \phi$. Since in the former case we know that a proposition $\phi$ ``satisfies'' $\bot$ whereas in the latter case a proposition $\phi$ ``implies'' $\bot$. Moreover, in the former case it is not a priori known that the proposition $\phi$ satisfies a falsum, it is also possible to derive something true, which is a posteriori a tautology, since then $(\phi \vdash \top)$ holds, and we have not obtained a contradiction w.r.t.~our premise. Thus, one starts with a proposition $\phi$ that is assumed to be true to establish if something inconsistent or consistent occurs w.r.t.~our premise to finally conclude that the premise $\phi$ is wrong or true. In contrast, for the latter case we know a priori, say due to $A \land \neg A = \bot$, that the proposition $\phi$ implies a falsum $\bot$, which also holds a posteriori. We get a wrong statement, since we know that $(\phi \Rightarrow \bot)$ is an invalid statement if $\phi$ is assumed to be true. Notice that the statement $(\phi \Rightarrow \bot)$ cannot be true. This follows from the assumption that $\phi$ is true, which implies that $\neg \phi$ must be false violating the equivalence of $(\phi \Rightarrow \bot) \Leftrightarrow \neg \phi$, consistency of the equivalence would require that $\neg \phi$ must be true, this cannot happen when $\phi$ is set to true. We deduce $(\phi \Rightarrow \bot)$ is a false statement if $\phi$ is assumed to be true. Note, the degree of freedom for $(\phi \Rightarrow \bot) \Leftrightarrow \neg \phi$ is one and not two as it is imposed by the authors. The premise $\phi$ implies a falsum but not a contradiction of our premise $\phi$. By equivalence, the false statement $\neg \phi$ determines that $(\phi \Rightarrow \bot)$ must be false too. Therefore the premise $\phi$ is true and not false as required. Thus, we do not observe a contradiction w.r.t.~$\phi$, but we observe a contradiction w.r.t.~a valid statement of $(\phi \Rightarrow \bot)$ if the premise $\phi$ is set to true. We realize that these authors have incorrectly applied $(\phi \Rightarrow \bot) \Leftrightarrow \neg \phi$. If we would follow the authors, we could always deduce that $\phi$ must be false, because a falsum occurs always. This means that we always get the desired result, and we could prove perverted results (see, for instance, Example~\ref{exp:elm}). Of course, this is a fallacy. 

The presented literature reflects only our research interest and should not be misunderstood as a representative survey. Moreover, we have chosen this sample according to the fact that these papers are irreversible flawed. Nevertheless, we guess that the described deficiencies are broader propagated as we might imagine. It is indispensable that the published results reflect a certain kind of reliability, otherwise we will observe in the literature contradictory results like Theorem $A$ and Theorem $\neg A$ are true, i.e., $(A \land \neg A) = \top$.      

The present paper is organized as follows: In the forthcoming section we introduce some notation and definitions applied in the discussed articles in order to make the presentation of the material more self-contained. Section~\ref{sec:indpr} discuss a first case from the field of the axiomatization of solution concepts. We quote the results and the essential parts of the authors argumentation followed by some reports of the committed logical mistakes. Whereas Section~\ref{sec:2exp} provides some further cases which are originated from the field of cooperative oligopoly games. We close our presentation with some final remarks in Section~\ref{sec:rem}.   
 
\section{Some Preliminaries}
\label{sec:prel}
In the sequel, we apply in essence the notation of the article~\citet{klep:13a}. For doing so, we let $U$ be a set, the universe of players, containing, without loss of generality, $1,\ldots,k$ whenever $\arrowvert U \arrowvert \ge k$. Here $\arrowvert U \arrowvert$ denotes the cardinality of $U$. A coalition is a finite nonempty subset of $U$. Let $\mathcal{F}$ denote the set of coalitions. A cooperative transferable utility game (TU game) is a pair $\langle N, v \rangle$ such that $N \in \mathcal{F}$ and $v: 2^{N} \rightarrow \mathbb{R}$ with $v(\emptyset):=0$. The real number $v(S) \in \mathbb{R}$ is called the value or worth of a coalition $S \in 2^{N}$. Let $S$ be a coalition, the number of members in $S$ will be denoted by $s:=|S|$. Let $\langle N, v \rangle$ be a TU game. We call $N$ its grand coalition and denote the set of all proper nonempty sub-coalitions of $N$ by $\mathcal{F}^{N}$, i.e. $\mathcal{F}^{N}= 2^{N}\backslash\{\emptyset,N\}$. Define respectively the set of feasible payoffs, the set of Pareto optimal feasible payoffs (pre-imputations), and the set of individually rational pre-imputations (imputations) of $\langle N, v \rangle$ by
\begin{equation*}
  \begin{split}
    X^{*}(N,v) & :=\{\mathbf{x} \in \mathbb{R}^{N} \,\arrowvert\, x(N) \le v(N)\},\\
    X(N,v) & := \{\mathbf{x} \in \mathbb{R}^{N} \,\arrowvert\, x(N) = v(N)\},\\
    I(N,v) & := \{\mathbf{x} \in X(N,v) \,\arrowvert\, x_{i} \ge v(\{i\}) \;\forall i \in N\}.
  \end{split}
\end{equation*}
where we apply $x(S) := \sum_{k \in S}\, x_{k}$ for every $S \in 2^{N}$, if $\mathbf{x} \in \mathbb{R}^{N}$, with $x(\emptyset):=0$. For $S \subset N$ and  $\mathbf{x} \in \mathbb{R}^{N}$, $\mathbf{x}_{S}$ denotes the restriction of $\mathbf{x}$ to $S$, i.e., $\mathbf{x}_{S}:= (x_{k})_{k \in S}$. Moreover, we identify a cooperative game by the vector $v := (v(S))_{S \subseteq N} \in \Gamma^{N} = \mathbb{R}^{2^{|N|}}$. In addition, we denote by $\Gamma_{I}$ the set of games $\langle N, v \rangle$ with $I(N,v)\neq\emptyset$, that is, $\langle N, v \rangle \in \Gamma_{I}$ iff $v(N) \ge \sum_{k}\,v(\{k\})$. 

A solution $\sigma$ assigns a subset $\sigma(N,v)$ of $X^{*}(N,v)$ to any game $\langle N, v \rangle$. Its restriction to a set $\Gamma$ of games is again denoted by $\sigma$. A solution on $\Gamma$ is the restriction to $\Gamma$ of a solution.

Given a vector $\mathbf{x} \in X(N,v)$, we define the {\bfseries excess} of coalition $S$ with respect to the pre-imputation $\mathbf{x}$ in the game $\langle N,v \rangle $ by 
\begin{equation} 
  \label{eq:exc} 
  e^{v}(S,\mathbf{x}):= v(S) - x(S). 
\end{equation} 

Take a game $v \in \Gamma^{N}$. For any pair of players $i,j \in N, i\neq j$, the {\bfseries maximum surplus} of player $i$ over player $j$ with respect to any pre-imputation $\mathbf{x} \in X(N,v)$ is given by the maximum excess at $\mathbf{x}$ over the set of coalitions containing player $i$ but not player $j$, thus\begin{equation} 
  \label{eq:maxexc} 
  s_{ij}(\mathbf{x},v):= \max_{S \in \mathcal{G}_{ij}} e^{v}(S,\mathbf{x}) \qquad\text{where}\;  \mathcal{G}_{ij}:= \{S \;\arrowvert\; i \in S\; \text{and}\; j \notin S \}. 
\end{equation} 
The set of all pre-imputations $\mathbf{x} \in X(N,v)$ that balances the maximum surpluses for each distinct pair of players $i,j \in N, i\neq j$ is called the~\hypertarget{hyp:prk}{{\bfseries pre-kernel}} of the game $v$, and is defined by 
  \begin{equation} 
    \label{eq:prek} 
    \mathcal{P\text{\itshape r}K}(N,v) := \left\{ \mathbf{x} \in X(N,v)\; \arrowvert\;  s_{ij}(\mathbf{x},v) = s_{ji}(\mathbf{x},v) \quad\text{for all}\; i,j \in N, i\neq j \right\}. 
  \end{equation} 

Related to the pre-kernel solution is the {\bfseries kernel} of a $n$-person game, which is the set of imputations $\mathbf{x} \in  I(N,v)$ satisfying for all $i,j \in N, i\neq j$
\begin{align}
 \label{eq:ker_sol}
  & \left[s_{ij}(\mathbf{x},v) - s_{ji}(\mathbf{x},v) \right]\cdot\left[x_{j}-v(\{j\})\right] \le 0 \quad\text{and}\\
  & \left[s_{ji}(\mathbf{x},v) - s_{ij}(\mathbf{x},v) \right]\cdot\left[x_{i}-v(\{i\})\right] \le 0.
\end{align}

In order to define the pre-nucleolus of a game $v \in \Gamma^{N}$, take any $\mathbf{x} \in \mathbb{R}^{N}$ to define a $2^{N}$-tuple vector $\theta(\mathbf{x})$ whose components are the excesses $e^{v}(S,\mathbf{x})$ of the $2^{N}$ coalitions $S \subseteq N$, arranged in decreasing order, that is,
\begin{equation}
 \label{eq:compl_vec}
  \theta_{i}(\mathbf{x}):=e^{v}(S_{i},\mathbf{x}) \ge e^{v}(S_{j},\mathbf{x}) =:\theta_{j}(\mathbf{x}) \qquad\text{if}\qquad 1 \le i \le j \le 2^{N}.
\end{equation}
Ordering the so-called complaint or dissatisfaction vectors $\theta(\mathbf{x})$ for all $\mathbf{x} \in \mathbb{R}^{N}$ by the lexicographic order  $\le_{L}$ on $\mathbb{R}^{N}$, we shall write
\begin{equation}
 \theta(\mathbf{x}) <_{L} \theta(\mathbf{y}) \qquad\text{if}\;\exists\;\text{an integer}\; 1 \le k \le 2^{N},
\end{equation}
such that $\theta_{i}(\mathbf{x}) = \theta_{i}(\mathbf{y})$ for $1 \le i < k$ and $\theta_{k}(\mathbf{x}) < \theta_{k}(\mathbf{y})$. Furthermore, we write $\theta(\mathbf{x}) \le_{L} \theta(\mathbf{y})$ if either $\theta(\mathbf{x}) <_{L} \theta(\mathbf{y})$ or $\theta(\mathbf{x}) = \theta(\mathbf{y})$. Now the {\bfseries pre-nucleolus} $\mathcal{P\text{\itshape r}N}(N,v)$ over the pre-imputations set $X(N,v)$ is defined by 
\begin{equation}
 \label{eq:prn_sol}
  \mathcal{P\text{\itshape r}N}(N,v) = \left\{\mathbf{x} \in X(N,v)\; \arrowvert\; \theta(\mathbf{x}) \le_{L} \theta(\mathbf{y}) \;\forall\; \mathbf{y} \in X(N,v) \right\}.
\end{equation}
The pre-nucleolus of any game $v \in \Gamma^{N}$ is non-empty as well as unique, and it is denoted as $\nu(N,v)$. Moreover, it is a sub-solution of the pre-kernel. In addition, notice that if the {\bfseries core} of a game $\langle\, N, v\,\rangle$ defined by
\begin{equation*}
  C(N,v):=\left\{ \mathbf{x} \in X(N,v)\; \arrowvert\, e^{v}(S,\mathbf{x}) \le 0\;\forall S \subseteq N \right\}
\end{equation*}
is non-empty, then the pre-nucleolus belongs to the core whenever the core is non-empty, that is, $\nu(N,v) \in C(N,v)$. 

Now the {\bfseries nucleolus} $\mathcal{N}(N,v)$ of a game $v \in \Gamma^{N}$ over the set $I(N,v)$ is defined as 
\begin{equation}
 \label{eq:nuc_sol}
  \mathcal{N}(N,v) := \left\{\mathbf{x} \in I(N,v)\; \arrowvert\; \theta(\mathbf{x}) \le_{L} \theta(\mathbf{y}) \;\forall\; \mathbf{y} \in I(N,v) \right\}.
\end{equation}
The set $\mathcal{N}(N,v)$ is a singleton whose unique element is referred to as $\nu_{I}(N,v)$. Similar to the pre-nucleolus, the nucleolus is a sub-solution of the kernel whenever the imputation set is non-empty. Moreover, if $C(N,v) \neq \emptyset$, then $\nu_{I}(N,v) \in C(N,v)$

Let us introduce the definition of a {\bfseries weighted (pre)-nucleolus}. A weight system is a system $\mathbf{p}:=(p^{N})_{N \in \mathcal{F}}$ such that for every $N \in \mathcal{F}$, $p^{N} := (p^{N}_{S})_{S \in \mathcal{F}^{N}}$, the weight system for $N$, satisfies $p^{N}_{S} > 0$ for all $S \in \mathcal{F}^{N}$. Let $\mathbf{p}$ be a weight system and $\langle N, v \rangle$ a TU game. The {\bfseries weighted pre-nucleolus} $\mathcal{P\text{\itshape r}N^{\mathbf{p}}}(N,v)$ and the {\bfseries  weighted nucleolus} $\mathcal{N^{\mathbf{p}}}(N,v)$ of $\langle N, v \rangle$ according to $\mathbf{p}$ are defined by 
\begin{equation*}
 \begin{split}
  \mathcal{P\text{\itshape r}N^{\mathbf{p}}}(N,v) & :=\mathcal{P\text{\itshape r}N}((p^{N}_{S}e^{v}(S,\cdot))_{S \in \mathcal{F}^{N}},X(N,v)),\\
  \mathcal{N^{\mathbf{p}}}(N,v) & :=\mathcal{N}((p^{N}_{S}e^{v}(S,\cdot))_{S \in \mathcal{F}^{N}},I(N,v)).
  \end{split}
\end{equation*}
Notice that also the set of the weighted pre-nucleolus $\mathcal{P\text{\itshape r}N^{\mathbf{p}}}(N,v)$ is a single point so that this unique element is referred to as $\nu^{\mathbf{p}}(N,v)$. Similar, for the set $\mathcal{N}^{\mathbf{p}}(N,v)$ which is a singleton and whose unique element is denoted as $\nu^{\mathbf{p}}_{I}(N,v)$.

Let $\mathbf{p}$ a weight system, $\langle N, v \rangle$ be a game, $\mathbf{x} \in \mathbb{R}^{N}$, and $i,j\in N, i \neq j $. The {\bfseries maximum $\mathbf{p}$-weighted surplus} of $k$ over $l$ at $\mathbf{x}$ w.r.t. $\langle N, v \rangle$ is defined by 
\begin{equation*}
    s^{\mathbf{p}}_{ij}(\mathbf{x},v):= \max_{S \in \mathcal{G}_{ij}} p^{N}_{S}\,e^{v}(S,\mathbf{x}) \qquad\text{where}\;  \mathcal{G}_{ij}:= \{S \;\arrowvert\; i \in S\; \text{and}\; j \notin S \}. 
\end{equation*}

The {\bfseries weighted pre-kernel} $\mathcal{P\text{\itshape r}K^{\mathbf{p}}}(N,v)$ and {\bfseries weighted kernel} $\mathcal{K^{\mathbf{p}}}(N,v)$ respectively, relative to the weight system $\mathbf{p}$ of a TU game $\langle N, v \rangle$ are defined by
  \begin{equation*} 
   \begin{split}
    \mathcal{P\text{\itshape r}K^{\mathbf{p}}}(N,v) & := \left\{ \mathbf{x} \in X(N,v)\; \arrowvert\;  s^{\mathbf{p}}_{ij}(\mathbf{x},v) = s^{\mathbf{p}}_{ji}(\mathbf{x},v) \quad\text{for all}\; i,j \in N, i\neq j \right\}, \\
    \mathcal{K^{\mathbf{p}}}(N,v) & := \left\{ \mathbf{x} \in I(N,v)\; \arrowvert\; s^{\mathbf{p}}_{ij}(\mathbf{x},v) \ge s^{\mathbf{p}}_{ji}(\mathbf{x},v) \;\text{or}\; x_{i}=v(\{i\}) \;\forall i,j \in N, i \neq j \right\}.
  \end{split}
  \end{equation*}  
Notice, that the weighted pre-nucleolus is an non-empty as well as unique solution which is a sub-solution of the weighted pre-kernel. Again, if the imputation set is non-empty, then the weighted nucleolus belongs to its weighted kernel. Moreover, if $C(N,v) \neq \emptyset$, then $\nu^{\mathbf{p}}(N,v) \in C(N,v)$. 

An {\bfseries objection} of player $i$ against a player $j$ w.r.t. a payoff vector $\mathbf{x} \in \mathbb{R}^{N}$ in game $v \in \Gamma^{N}$ is a pair $(\mathbf{y}_{S},S)$ with $S \in \mathcal{G}_{ij}$ and $\mathbf{y}_{S} := \{y_{k}\}_{k \in S}$ satisfying the following properties:
\begin{equation}
  \label{eq:objec}
  v(S) = \sum_{k \in S}\, y_{k} \qquad\text{and}\qquad y_{k} > x_{k} \quad\text{for}\; k \in S.
\end{equation}
A {\bfseries counter-objection} to the objection $(\mathbf{y}_{S},S)$ is a pair $(\mathbf{z}_{T},T)$ with $T \in \mathcal{G}_{ji}$ and $\mathbf{z}_{T}:=\{z_{k}\}_{k \in T} $ satisfying 
\begin{equation}
  \label{eq:cobjec}
  \begin{split}
  v(T) = \sum_{k \in T}\, z_{k} \qquad\text{and}\qquad z_{k} & \ge x_{k} \quad\text{for}\; k \in T \backslash S \\
  z_{k} & \ge y_{k} \quad\text{for}\; k \in T \cap S.
  \end{split}
\end{equation}
Thus, if the pair $(\mathbf{y}_{S},S)$ is an objection against vector $\mathbf{x}$, then any member of coalition $S \in \mathcal{G}_{ij}$ can improve upon rather than accepting proposal $\mathbf{x}$. Acceptance would mean that players in $S  \in \mathcal{G}_{ij}$ would accept a loss due to $e^{v}(S,\mathbf{x})>0$. Hence, a player $i$ can formulate an objection against player $j$ using coalition $S \in \mathcal{G}_{ij}$ w.r.t. the proposal $\mathbf{x}$ iff the excess $e^{v}(S,\mathbf{x})$ is positive.

In contrast, a counter-objection $(\mathbf{z}_{T},T)$ of player $j$ against player $i$ w.r.t. objection $(\mathbf{y}_{S},S)$ uses a coalition $T$ without player $i$, i.e. $T \in \mathcal{G}_{ji}$, to formulate a proposal that cannot strictly be improved upon to the precedent proposal for players belonging to the set $S \cap T$ and which can also not strictly be improved upon w.r.t. $\mathbf{x}$ for all $k \in T\backslash S$. This means, that player $j$ can only use a coalition $T \in \mathcal{G}_{ji}$ with non-negative excess $e^{v}(T,\mathbf{x})$ to formulate a counter-objection against player $i$.

An imputation $\mathbf{x} \in I(N,v)$ is an element of the {\bfseries bargaining set} $\mathcal{M}(N,v)$ of game $v \in \Gamma^{N}$ whenever for any objection of a player against another player w.r.t. $\mathbf{x}$ in $v \in \Gamma^{N}$ exists a counter-objection. The bargaining set can be empty whenever the imputation set is empty. For zero-normalized games the imputation set is never empty, and therefore the bargaining set $\mathcal{M}(N,v)$ exists, which contains the nucleolus and kernel of the game, i.e. $\nu_{I}(N,v) \subseteq \mathcal{K}(N,v) \subseteq \mathcal{M}(N,v)$. 

Let $\sigma$ be a solution on a set $\Gamma$ of games. A solution $\sigma$ may satisfy some of the following possible properties: 

\begin{labeling}[:]{Axioms}
\item[\bfseries{Non-Emptiness (NE)}] If $\sigma(N,v) \not=\emptyset$ for all $\langle N, v \rangle \in \Gamma$.
\item[\bfseries{Single-Valuedness (SIVA)}] If $\arrowvert\,\sigma(N,v)\, \arrowvert = 1$ for every $\langle\, N, v\,\rangle  \in \Gamma$.
\item[\bfseries{Pareto-Optimality (PO)}] If $\sigma(N,v) \in X(N,v)$ for all $\langle N, v \rangle \in \Gamma$.
\item[\bfseries{Anonymity (AN)}] If for $\langle\, N, v\,\rangle  \in \Gamma$, for an injection $\pi: N \rightarrow \mathcal{U}$ and for $\langle\, \pi(N), \pi v\,\rangle \in \Gamma$ implying $\sigma(\pi(N), \pi v) = \pi(\sigma(N,v))$.
\item[\bfseries{Symmetry (SYM)}] If $\sigma(N,v) = \pi(\sigma(N,v))$ for all $\langle N, v \rangle \in \Gamma$ and all symmetries $\pi$ of $\langle N, v \rangle$.
\item[\bfseries{Individual Rationality (IR)}] If $\langle\, N, v\,\rangle \in \Gamma$ and $\vec{x} \in \sigma(N,v)$, then $x_{k} \ge v(\{k\})$ for all $k \in N$. 
\item[\bfseries{Equal Treatment Property (ETP)}] If $\langle\, N, v\,\rangle \in \Gamma$, $\vec{x} \in \sigma(N,v)$ and if $k$ and $l$ are substitutes, i.e., $v(S \cup \{k\}) = v(S \cup \{l\})$ for all $S \subseteq N\backslash\{k,l\}$, then $x_{k} = x_{l}$. 
\item[\bfseries{Covariance with Strategic Equivalence (COV)}] If for $\langle\, N, v_{1}\,\rangle, \langle\, N, v_{2}\,\rangle \in \Gamma$, with $v_{2} = t \cdot v_{1} + \mathfrak{m} $ for some $t \in \mathbb{R}_{++}, \mathfrak{m} \in \mathbb{R}^{2^{N}}$, then $\sigma(N,v_{2}) = t \cdot \sigma(N,v_{1}) + \mathbf{m}$, whereas $\mathbf{m} \in \mathbb{R}^{N}$ and $\mathfrak{m}$ is the vector of measures obtained from $\mathbf{m}$.
\end{labeling}

\section{The Case of the Indirect Proof}
\label{sec:indpr}

We quote now some statements from~\citet{klep:13a} and discuss their proofs in order to observe how deficient these authors have applied the indirect proof. The essential arguments and conclusions of the authors are set in italic and are highlighted by a red coloring.   

\begin{quote}
\begin{labeling}[:]{Theorem}
\item[\bfseries{Theorem 3.3~(\citet[p. 7]{klep:13a})}]
\itshape{Let $\mathbf{p}$ be a weight system, $\Gamma \supseteq \Gamma_{I}$, and $\sigma$ be one of the following solutions on $\Gamma$: $\mathcal{N^{\mathbf{p}}}, \mathcal{P\text{\itshape r}N^{\mathbf{p}}},\mathcal{K^{\mathbf{p}}}$ or $\mathcal{P\text{\itshape r}K^{\mathbf{p}}}$. Then $\sigma$ satisfies {\bfseries ETP} if and only if $\mathbf{p}$ is symmetric.}
\end{labeling}
\begin{proof}
  The ``if-part'' is an obvious consequence of the definitions of the considered weighted solutions. In order to show the ``only-if-part'' let $\sigma$ be one of the considered solutions and {\itshape \color{red} let it satisfy ETP}. Assume, on the contrary, that {\itshape \color{red} $\mathbf{p}$ does not satisfy the desired property}. Hence, there exists a coalition $N$ and some $S,S^{\prime} \in \mathcal{F}^{N}$ with $\arrowvert S \arrowvert = \arrowvert S^{\prime} \arrowvert $ such that $p^{N}_{S} \neq p^{N}_{S^{\prime}}$. {\itshape \color{red} It remains to show that $\sigma$ violates ETP}. As $S^{\prime}$ arises from $S$ by a sequence of replacements of one player by one other player, we may assume that $\arrowvert S\backslash S^{\prime} \arrowvert = 1$. Let $T,k,l$ be determined by $S=T\cup \{k\}$ and $S^{\prime} = T \cup \{l\}$. Let $\langle N, v \rangle$ be the game defined by $v(N)=v(T)=v(N\backslash T) = 0, v(T\cup \{i\}) = -1$ for all $i \in N\backslash T $, and $v(R)=\frac{-p^{N}_{S} - p^{N}_{S^{\prime}}}{\min\{p^{N}_{Q} \,\arrowvert\, Q \in \mathcal{F}^{N}\}}$ for all other $R \in \mathcal{F}^{N}$. Then $\langle N, v \rangle \in \Gamma_{I}$. Let $\mathbf{y} = \nu^{\mathbf{p}}(N,v)$. By Remark 2.4(1), $\mathbf{y} \in \sigma(N,v)$. {\itshape \color{red} As $\sigma$ satisfies ETP} and as all players inside $T$ are substitutes and all players in $N\backslash T$ are substitutes as well, there exist $\alpha,\beta \in \mathbb{R}$ such that $y_{i}= \alpha$ for all $i \in T$ and $y_{j}=\beta$ for all $j \in N\backslash T$. As $y(N) = v(N) = 0, \arrowvert T \arrowvert \, \alpha + \arrowvert N\backslash T \arrowvert \, \beta = 0$. Let $\mathbf{x} = \mathbf{0} \in \mathbb{R}^{N}$. Then $e^{v}(T,\mathbf{x}) = e^{v}(N\backslash T,\mathbf{x}) = 0$ and $e^{v}(R,\mathbf{x}) < 0$ for all $R \in \mathcal{F}^{N}\backslash\{T,N\backslash T\}$. By the definition of the weighted pre-nucleolus, $e^{v}(T,\mathbf{y}) = e^{v}(N\backslash T,\mathbf{y}) = 0$. Hence, $y(T)=y(N\backslash T)=0$ implying $\arrowvert T \arrowvert \,\alpha = \beta = 0$, i.e., $\mathbf{y}=\mathbf{x}$.
 For any $R \in \mathcal{F}^{N}\backslash\{S\}$ with $k \in R \not\ni l$, the definition of $v$ gives
 \begin{equation*}
   p^{N}_{R}\,e^{v}(R,\mathbf{y}) \le p^{N}_{R}\,\frac{-p^{N}_{S} - p^{N}_{S^{\prime}}}{p^{N}_{R}} < -p^{N}_{S} = p^{N}_{S}\,e^{v}(S,\mathbf{y}). 
 \end{equation*}
A similar argument is valid when switching the roles of $k$ and $l$, so $s^{\mathbf{p}}_{kl}(\mathbf{y},v) = -p^{N}_{S} \neq -p^{N}_{S^{\prime}} = s^{\mathbf{p}}_{lk}(\mathbf{y},v) $. {\itshape \color{red} Hence, $\mathbf{y} \notin \mathcal{P\text{\itshape r}K^{\mathbf{p}}}(N,v), \mathbf{y} \notin \mathcal{K^{\mathbf{p}}}(N,v)$} and {\itshape \color{red} the desired contradiction is obtained} by Remark 2.4(1).~(\citet[pp. 7-8]{klep:13a})
\end{proof}
\end{quote}
We give now the reasons why Theorem 3.3 cannot be correct. For this purpose, we introduce two truth tables. A logical statement/proposition is formed by the symbols $A$ or $B$, which means that a statement $A$ is true or false. However, the inversion is formed by the negation of a proposition by using the logical term ``not'' denoted by $\neg$. If $A$ is a proposition, then $\neg A$ is the negation of $A$ verbalized as ``not $A$'' or ``$A$ is false''. The effect of negation, conjunction, disjunction, and implication on the truth values of logical statements is summarized by a so-called truth table. In this table, the capital letter {\bfseries T} indicates a true proposition and {\bfseries F} indicates that it is false. 

\begin{center}
\begin{tabular}{c c c c c c c c c c c}
\hline
$A$ & $B$ & $\neg B$ & $A \Rightarrow B$ & $ \neg (A \Rightarrow B)$ & $A \Leftarrow B$ & $A \Leftrightarrow B$ & $A \lor \neg B$ & $A \land B $ &  $A \lor B $\\
\hline
F & F & T & T & F & T & T & T & F & F \\
F & T & F & T & F & F & F & F & F & T \\
T & F & T & F & T & T & F & T & F & T \\
T & T & F & T & F & T & T & T & T & T \\ %\hline
\end{tabular}

\begin{tabular}{c c c c |c c ||c c| c c}
\hline
$A$ & $B$ &  $\neg A$ & $\neg B$ & $\neg A \Rightarrow \neg B$ & $A \lor \neg B$ & $\neg A \Leftarrow \neg B$ & $\neg A \lor B$  & $A \land \neg B$ & $\neg A \Leftrightarrow \neg B$ \\
\hline
F & F & T & T & T & T & T & T & F & T   \\
F & T & T & F & F & F & T & T & F & F   \\
T & F & F & T & T & T & F & F & T & F   \\
T & T & F & F & T & T & T & T & F & T   \\\hline
\end{tabular}
\end{center}
Two statements are indicated as logically equivalent through the symbol $\equiv$. For instance, by the truth table we realize that the two statements $\neg A \Leftarrow \neg B$ and $\neg A \lor B$ are logically equivalent, which is formally expressed by $(\neg A \Leftarrow \neg B) \equiv (\neg A \lor B)$. A falsum $\bot$ is, for instance, the conjunction $A \land \neg A$ whereas a tautology $\top$ can be expressed, for instance, by the disjunction $\neg A \lor A$. Moreover, a proposition or premise $A$ might satisfy a falsum or a tautology or an arbitrary property $B$, which is expressed by $(A \vdash \bot)$ or $(A \vdash \top)$ or $(A \vdash B)$ respectively. This should not be confounded with an implication of the form $(A \Rightarrow \bot)$ or $(A \Rightarrow \top)$ or $(A \Rightarrow B)$ respectively.    

\noindent In Theorem 3.3,~\citeauthor{klep:13a} claim that $\sigma$ (weighted (pre-)nucleolus/(pre-)kernel) satisfies

{\bfseries ETP} ($A$ is true) if, and only if, the weight system $\mathbf{p}$ is symmetric ($B$ is true).

\noindent The {\bfseries contrapositive} of the Theorem states that $\sigma$ fulfills

not {\bfseries ETP} ($\neg A$) if, and only if, the weight system $\mathbf{p}$ is asymmetric ($\neg B$).

\noindent \citeauthor{klep:13a} discuss the ``only if part'', i.e., if $\sigma$ satisfies {\bfseries ETP} ($A$ is true), then the weight system $\mathbf{p}$ is symmetric ($B$ is true). They apply their indirect proof with elements of a {\bfseries material implication}. A material implication is a rule of replacement that allows to replace a conditional proposition by a disjunction. For instance, the conditional statement $A$ implies $B$ can be replaced by the disjunction $\neg A \lor B$, which is logically equivalent to the former proposition (see the truth table). In contrast, an {\bfseries indirect proof} is based on the fact that either a logical statement is true or false but not both. This proof technique is also known under the name ``reductio ad absurdum'', i.e., one leads an ``argumentum ad absurdum'' or to a ``reduction to absurdity''. This is a common form of argument seeking to demonstrate that a statement is true by showing that a false, untenable, or absurd result follows from  its denial, or in turn to demonstrate that a proposition is false by showing that a false, untenable, or absurd result follows from its acceptance. Formally, a proof by contradiction tries to establish $(\phi \vdash \bot) \Leftrightarrow \neg \phi$, this should not be confounded with $(\phi \Rightarrow \bot) \Leftrightarrow \neg \phi$. Doing so, can provoke severe fallacy, this will be more thoroughly discussed in sequel.  

In this case,~\citeauthor{klep:13a} want to prove that whenever $A$ is true, then $B$ is also valid, which is equivalent to if $\neg B$ then $\neg A$. Moreover, from the above truth table we observe that if $\neg B$, then $\neg A$ is equivalent to $\neg A \lor B$, but not to $A \land \neg B$. By the truth table, it should be evident that the conjunction $A \land \neg B$ is the negation of the disjunction $\neg A \lor B$, and it is not its contrapositive. Obviously, $A \land \neg B$ is logically equivalent to $\neg (A \Rightarrow B)$, which is not equivalent to $\neg B \Rightarrow \neg A$. If $A \Rightarrow B$ then we can focus on the negation $\neg (A \Rightarrow B) \equiv \neg (\neg A \lor B) \equiv (A \land \neg B) $, since $A \Rightarrow B$ is logically equivalent to $A \land \neg B \Rightarrow B \land \neg B$, which imposes a proof by contradiction. Similar, if we have the contrapositive $\neg B \Rightarrow \neg A$, we can prove this by an indirect proof through if $A \land \neg B \Rightarrow A \land \neg A$. This allows one to infer that $A \Rightarrow B$ is valid or invalid. 

For their proof of Theorem of 3.3, they try to run an indirect proof while relying on a material implication.\footnote{It is inconceivable for us that the considered articles are based on a circular reasoning (circulus in probando), for this reason we focus in the sequel on the indirect proof based on a material implication. Obviously, there is only a slight change in the argumentation necessary to incorporate in our discussion the circular argument from the introduction. This means that the considered case imposes no loss of generality on our line of argument, which implies that in both cases the authors disprove themselves.} For doing so, they assume that $A$ and $\neg B$ is fulfilled in order to get a contradiction, because the conjunction $A \land \neg B$ is the negation of the disjunction $\neg A \lor B$. If they have obtained their contradiction, they assume that the proposition $\neg A \lor B$ is true, to finally infer that the implication $A \Rightarrow B$ is true as well. But this is not a permissible implementation, since one cannot suppose first that $A \land \neg B$ is given to conclude that $\neg A \lor B$ is valid or invalid, that is based on the preceding truth or falsehood of a statement. By an indirect proof, the conjunction $A \land \neg B$ implies something false, that is, one introduces a prerequisite $A \land \neg B$ that is assumed to be true, and yields the implication to a falsehood, for instance, that $A \land \neg A$ is invalid. Then, we know that the implication $A \land \neg B \Rightarrow A \land \neg A$ is a wrong proposition. As a consequence, the implication $A \Rightarrow B$ is invalid as well, due to $(A \land \neg B \Rightarrow A \land \neg A) \equiv (A \Rightarrow B)$. However, if $A \land \neg B$ is assumed to be false, then $A \land \neg A$ is invalid too. The proposition is a valid outcome, the implication $A \Rightarrow B$ is valid as well. This, and only this, is the correct line of argument.  

\begin{example}
  \label{exp:mimpl}
Let us look at a statement $A$ like ``he is a game theorist'' and $\neg B$ ``he has not mastered propositional logic''. Furthermore, consider three propositions: 
\begin{labeling}[:]{disjunction}
 \item[\bfseries{Implication $A \Rightarrow B$}] if ``he is a game theorist'', then ``he has mastered propositional logic''.
 \item[\bfseries{Contrapositive $\neg B \Rightarrow \neg A$}] if ``he has not mastered propositional logic'', then ``he is not a game theorist''.
 \item[\bfseries{Disjunction $\neg A \lor B$}] ``he is not a game theorist'' or ``he has mastered propositional logic''.
\end{labeling}
If the outcome of some logical inference is ``he is a game theorist'' and ``he has not mastered propositional logic'', that is, $A \land \neg B$ is valid, then all three statements are false. However, if $A \land \neg B$ is invalid, then all statements are true. This can be accomplished by applying an indirect proof. Thus, we do not assume for the latter case that ``he is a game theorist'' and ``he has not mastered propositional logic'' ($A \land \neg B$ is invalid) to derive some false statement like ``he is a game theorist'' and ``he is not a game theorist'' ($A \land \neg A$), this means, that the whole proposition is a truth, we infer from this outcome that all three statements must be satisfied. However, if it is given that ``he is a game theorist'' and ``he has not mastered propositional logic'' ($A \land \neg B$ is valid), and we get the false proposition ``he is a game theorist'' and ``he is not a game theorist'' ($A \land \neg A$), we infer that this invalids the implication. Therefore, all three statements are wrong. 
\end{example}

By an indirect proof,~\citeauthor{klep:13a} have to establish that whenever it is false that the solution $\sigma$ satisfies the conjunction {\bfseries ETP} ($A$ is valid) and $\mathbf{p}$ is asymmetric ($\neg B$), then {\bfseries ETP} ($A$) and non {\bfseries ETP} ($\neg A$) are a falsehood on $\sigma$. Hence, the proposition {\itshape if $\sigma$ satisfies {\bfseries ETP} ($A$ is true), then the weight system $\mathbf{p}$ is symmetric ($B$ is true)} is a truth, since a false statement implies something false. This means that for an indirect proof, one starts with a claim that is assumed to be false and leads this claim to a contradiction. Then, one can infer that the proposition, that should be proved, is a truth. 

In contrast, they start with let $\sigma$ ``{\itshape satisfy ETP}'', i.e., $A$ is true, and then supposing in the next step that $\mathbf{p}$ is asymmetric ($\neg B$), in order to construct a game from which they try to derive a contradiction. This means, they assume that $\sigma$ satisfies {\bfseries ETP} and the weight system $\mathbf{p}$ is asymmetric, from which they want to show that a contradiction can be drawn, that is, something false follows. However, by the above consideration, it should be evident this is mystified, and is therefore a fallacy. Nevertheless, we have to observe by their proof that this prerequisite will be used in the sequel by their phrase ``{\itshape as $\sigma$ satisfies ETP}'' to finally derive that $\mathbf{y} \notin \mathcal{P\text{\itshape r}K^{\mathbf{p}}}(N,v), \mathbf{y} \notin \mathcal{K^{\mathbf{p}}}(N,v)$ follow, which is their ``{\itshape desired contradiction}'' that $\sigma$ does not satisfy {\bfseries ETP} ($\neg A$). In effect, they have disproved their Theorem 3.3, because they have shown that a true prerequisite implies a wrong claim, however, this implication is a wrong statement. As a consequence, we conclude that the implication $A \land \neg B \Rightarrow A \land \neg A$ is wrong, and in accordance with $(A \land \neg B \Rightarrow A \land \neg A) \equiv (A \Rightarrow B)$, we get that $A \Rightarrow B$ must be false either. Hence, the proposition {\itshape if $\sigma$ satisfies {\bfseries ETP} ($A$ is true), then the weight system $\mathbf{p}$ is symmetric ($B$ is true)} is a falsehood. They disproved themselves, since the authors have shown the exact opposite of what had been intended. We infer from that, Theorem 3.3 is false.

To see that from a false conclusion a false implication follows, can be observed from an example taken from an elementary course in mathematics. 

\begin{example}
 \label{exp:elm}
Let $m$ denote an arbitrary number, and let us ``{\itshape prove}'' the wrong implication that

if $m^2$ is even ($A$), then $m$ is odd ($B$),

\noindent while running a purported proof by the arguments used by~\citeauthor{klep:13a}. In a first step, we assume that $A \land \neg B$ is valid. For this purpose, we suppose that $m$ is even ($\neg B$) s.t. $m=2\;k$ for some integer $k$, and assume that $m^2$ is even too ($A$ is true), i.e., $m^2=2\;q$ for some integer $q$, then we get that $m^2=(2\;k)^2=4\;k^2 = 2\;q$. This implies $k=\pm\,\sqrt(q/2)$, which is the {\itshape desired contradiction}. We conclude that $m$ is odd ($B$). Hence, a valid premise $A \land \neg B$ implies something wrong ($B \land \neg B$), which is a true proposition by~\citeauthor{klep:13a}. Therefore, these authors would conclude that $A \land \neg B$ is wrong, then the negation of this expression, i.e., $\neg A \lor B$ is true. From which they would deduce that $A \Rightarrow B$ is a valid statement. This is certainly a fallacy, one incorrectly applied $(\phi \Rightarrow \bot) \Leftrightarrow \neg \phi$. However, it should be obvious by the preceding discussion that this gives in fact a disproof of $A \Rightarrow B$, thus we have $A \not\Rightarrow B$.
\end{example}

By the consideration from above, we realize that~\citeauthor{klep:13a} have shown that a valid premise $A \land \neg B$ implies a falsehood, which is a wrong statement. Remember that the implications $A \land \neg B \Rightarrow A \land \neg A$ or $A \land \neg B \Rightarrow B \land \neg B$ are logically equivalent to $A \Rightarrow B$. Hence, if one has shown that such an implication or every other implication that should be equivalent to $A \Rightarrow B$ produces a wrong proposition, one has to conclude that $A \Rightarrow B$ must be invalid too. In this case, on cannot deduce that $A \land \neg B$ is false, this is due that $A \land \neg B$ was assumed to be valid. Applying then that $A \land \neg B$ is false in order to infer from this, that its negation $A \lor \neg B$ as well as the implication $A \Rightarrow B$ must be valid, is, of course, a fallacy. 

Now, we shall give some arguments of how the proof must run to get the desired logical proposition. This will also demonstrate that Theorem 3.3 cannot be saved, and therefore the whole article is false.  

\citeauthor{klep:13a} have to show that the weighted pre-nucleolus $\mathbf{y} = \nu^{\mathbf{p}}(N,v)$ is unequal to the null-vector. By the construction of the game, the players $k$ and $l$ are substitutes, from that $\mathbf{y} \not= \mathbf{0}$ must follow. Then, they have to show that $y_k \not= y_l$ such that $y_k=-\beta$ and $y_l=\beta$ is given. Such a result can be now deduced from the constructed game, since {\bfseries ETP} is not anymore assumed (see also Example~\ref{exp:netp}). Hence, {\bfseries ETP} is false ($\neg A$). This would have been the final step by a proof by contraposition, i.e., $\neg B \Rightarrow \neg A$. From Example 3.6 below, that gives an unintended counter-example by the authors, we can even learn that such a result cannot be guaranteed.

In the next step, we observe by following the arguments of~\citeauthor{klep:13a} for their proof of Proposition 3.5 that they repeat this fallacy. They are again confused between the propositional statements of a proof by contradiction and the material implication. We do not want to bother the readership while representing their whole lengthy proof of Proposition 3.5, we, therefore, confine ourselves on the main faulty arguments applied by the authors.

\begin{quote}
\begin{labeling}[:]{proposition}
\item[\bfseries{Proposition 3.5~(\citet[p. 8]{klep:13a})}]
  If $\mathbf{p}$ is a symmetric weight system, then for any game $\langle N,v \rangle$, $\mathcal{P\text{\itshape r}K^{\mathbf{p}}}(N,v)$ is compact.
\end{labeling}
\begin{proof}
Assume, {\itshape \color{red} on the contrary, that $\mathcal{P\text{\itshape r}K^{\mathbf{p}}}(N,v)$ is not compact}. Let $\mathcal{S}=(S^{kl})_{k,l \in N \times N,k \neq l}$ be a constellation such that $X_{S}$ is unbounded. Let $(\mathbf{x}^{r})_{r \in \mathbb{N}}$ be an unbounded sequence of elements of $X_{S}$. Then, after replacing (\ldots) {\itshape \color{red} Since $\mathbf{p}$ is symmetric}, $p^{N}_{R}\,e^{v}(R,\mathbf{x}^{r}) > p^{N}_{S^{lk}}\,e^{v}(S^{lk},\mathbf{x}^{r}) = p^{N}_{S^{kl}}\,e^{v}(S^{kl},\mathbf{x}^{r}) = \mu^{r}$ for $r$ taken sufficiently large, so the desired contradiction has been obtained.~(\citet[p. 8]{klep:13a})
\end{proof}
\end{quote}
The authors have to show by the proposition that 

if the weight system $\mathbf{p}$ is symmetric ($A$ is true), then for any game, $\mathcal{P\text{\itshape r}K^{\mathbf{p}}}(N,v)$ is compact ($B$ is true).

\noindent For a proof by contraposition, they have to establish that the equivalent argument  

if for any game, $\mathcal{P\text{\itshape r}K^{\mathbf{p}}}(N,v)$ is not compact ($\neg B$), then the weight system $\mathbf{p}$ is asymmetric ($\neg A$),

\noindent holds true.

The authors start by the assumption that $\mathcal{P\text{\itshape r}K^{\mathbf{p}}}(N,v)$ is not compact ($\neg B$), and select a sequence which is unbounded to derive a contradiction. Then again, they assume that $\mathbf{p}$ is symmetric ($A$ is true) to get a so-called desired contradiction. To summarize, they introduce a valid premise $A \land \neg B$ to obtain a contradiction. By the same reasoning as above, this argument is misguided. One cannot conclude, whenever something is true from which a false implication follows, that this a true proposition. Again, they have to show that whenever $A \land \neg B$ is invalid, a wrong claim will be obtained, i.e., a contradiction follows in order to infer that the conclusion $A \Rightarrow B$ can be drawn. Once more,~\citeauthor{klep:13a} have disproved their own Proposition 3.5. We conclude their proposition is wrong as well. A further component of invalidating their results.  

\noindent Next, let us consider the unintended counter-example of~\citeauthor{klep:13a} to Theorem 3.3.

\begin{quote}
\begin{labeling}[:]{example}
\item[\bfseries{Example 3.6~(\citet[p. 9]{klep:13a})}]
Let $N=\{1,\ldots, 5\}$ and $p^{N}$ be defined by
\begin{equation*}
  p^{N}_{S} = 7 \qquad \text{if}\; \arrowvert S \cap \{1,2,3\}\arrowvert = 2 \;\text{and}\; \arrowvert S \cap \{4,5\} \arrowvert = 1 \;\text{and}\; p^{N}_{S} = 1\; \text{otherwise},
\end{equation*}
for all $S \in \mathcal{F}^{N}$. Then $\mathbf{x}^{t} = (-2t,-2t,-2t,3t,3t) \in \mathcal{P\text{\itshape r}K^{\mathbf{p}}}(N,\mathbf{0})$ for all $t \ge 0$. Indeed, the maximal $\mathbf{p}$-weighted excess at $\mathbf{x}^{t}$ is attained by the coalition $S$ with $p^{N}_{S} = 7$, and it is $7t$. However, the set of these coalitions is completely separating, i.e., for any $k,l \in N, k \neq l$, there exists a coalition $S \in \mathcal{F}^{N}$ with $p^{N}_{S} = 7$ and $l \notin S \ni k$ so that $s^{\mathbf{p}}_{kl}(\mathbf{x}^{t},v) = 7t$. Hence, this weighted pre-kernel is unbounded.
\end{labeling}
\end{quote}
Example 3.6 demonstrates for an asymmetric weight system $\mathbf{p}$, and for $v=\mathbf{0}$, that the derived weighted pre-kernel is not compact. The example is correct related to the proposition of the weighted pre-kernel. However, in contrast to their proof for Theorem 3.3, we have $s^{\mathbf{p}}_{kl}(\mathbf{x}^{t},v) = 7t > p^{N}_{S} = p^{N}_{S^{\prime}} = 3 t$ whenever $t >0$ for $S = \{1,2,3,4\}$ and $S^{\prime} = \{1,2,3,5\}$. By the above discussion, it should, however, be evident that even Proposition 3.5 is false due to the fact that~\citeauthor{klep:13a} make the same wrong conclusion as in their proof of Theorem 3.3. Moreover, they apply in their example an ambiguous argument. They introduce an asymmetric weight system and obtain after some manipulation the result that the weighted per-kernel is not a compact solution set. Thus, they have discussed an example where the introduced weight system $\mathbf{p}$ is asymmetric ($\neg A$), and as a consequence, the derived weighted pre-kernel solution is non-compact ($\neg B$). Reading the statement of their Proposition 3.5, we realize, however, that they must demonstrate by their example the reverse statement that whenever the weighted pre-kernel is non-compact ($\neg B$), then the weight system $\mathbf{p}$ must be asymmetric ($\neg A$). Both propositions are logically not equivalent. Thus, we can again conclude that the observed non-compactness is not obtained by the asymmetric weight system $\mathbf{p}$. It makes even not so much sense to us. Nevertheless, we have $v=\mathbf{0}$, all players are substitutes. Note that the weighted pre-kernel contains the weighted pre-nucleolus, which is here the null-vector\footnote{Confirmed from Peter Sudh\"olter by private conversation.}. But, if the weight system $\mathbf{p}$ is asymmetric, the weighted pre-nucleolus cannot be given by the null-vector due to Theorem 3.3. On the contrary, the weighted pre-nucleolus distributes the null-vector, and satisfies therefore {\bfseries ETP}, invalidating Theorem 3.3, and as a consequence their results. We observe that this example confirms the disproof of Theorem 3.3 by~\citeauthor{klep:13a}.

We discuss now another counter-example where the weighted pre-kernel coincides with the weighted pre-nucleolus while distributing the null-vector. 

\begin{example}
 \label{exp:etp} 
Let $p^{N}$ be defined as by Example 3.6 from~\citet{klep:13a}, hence the weight system $\mathbf{p}$ is asymmetric. Define next the TU game as in their proof of Theorem 3.3 from~\citet{klep:13a}, that is, the game is defined by $v(N)=v(T)=v(N\backslash T) = 0, v(T\cup \{i\}) = -1$ for all $i \in N\backslash T $, and $v(R)=\frac{-p^{N}_{S} - p^{N}_{S^{\prime}}}{\min\{p^{N}_{Q} \,\arrowvert\, Q \in \mathcal{F}^{N}\}}$ for all other $R \in \mathcal{F}^{N}$. Here, coalition $T$ is given by $\{1,2,3\}$, and the complement of coalition $T$ by $\{4,5\}$. Choose $k=4 \neq l=5$, coalition $S$ is determined by $T\cup \{k\}$ and $S^{\prime}$ by $T \cup \{l\}$. Then, we obtain an asymmetric TU game given by
\begin{equation*}
  v(N)=v(\{1,2,3\}) = v(\{4,5\}) = 0, \quad v(\{1,2,3,4\}) = v(\{1,2,3,5\}) = -1, \quad v(R) = -2,
\end{equation*}
for all other $R \in \mathcal{F}^{N}$. Recall that the weight system $\mathbf{p}$ is asymmetric, whereas the unique weighted pre-kernel coincides with the weighted pre-nucleolus, which is the null-vector. This result violates the outcome of their proof of Theorem 3.3 that $\nu^{\mathbf{p}}(N,v) \neq \mathbf{0}$ should hold. Again, the weight system $\mathbf{p}$ is asymmetric, and the weighted pre-nucleolus as well as the weighted pre-kernel satisfy {\bfseries ETP}. Here, players $\{1,2,3\}$ and $\{4,5\}$ are substitutes. Contradicting the fact that according to Theorem 3.3 of~\citet{klep:13a} the weighted pre-kernel should not satisfy {\bfseries ETP}. We discussed a further example for their disproof of Theorem 3.3.

\end{example}

The next example demonstrates that wrong conclusions are drawn for their proof of Theorem 3.3 by~\citeauthor{klep:13a}, when we impose {\bfseries ETP} as an assumption rather than a result from logical deduction.

\begin{example}
  \label{exp:netp}
Let $N=\{1,2,3,4\}$ and $p^{N}$ be defined by
\begin{equation*}
  p^{N}_{S} = 3 \qquad \text{if}\; \arrowvert S \cap \{1,3\}\arrowvert = 2 \;\text{and}\; \arrowvert S \cap \{2,4\} \arrowvert = 1 \;\text{and}\; p^{N}_{S} = 1\; \text{otherwise},
\end{equation*}
for all $S \in \mathcal{F}^{N}$. Hence, the weight system $\mathbf{p}$ is asymmetric. Define next the TU game as in their proof of Theorem 3.3 from~\citet{klep:13a}. Let $T=\{1,2\},k=3,l=4$, then the complement of coalition $T$ is determined by $\{3,4\}$. Moreover, coalition $S$ is given by $T \cup \{k\} = \{1,2,3\}$ and coalition $S^{\prime}$ by $T \cup \{l\} = \{1,2,4\}$. Then, we obtain an asymmetric TU game that is quantified by
\begin{equation*}
  v(N)=v(\{1,2\}) = v(\{3,4\}) = 0, \quad v(\{1,2,3\}) = v(\{1,2,4\}) = -1, \quad v(R) = -4,
\end{equation*}
for all other $R \in \mathcal{F}^{N}$. In this game, players $\{1,2\}$ are substitutes as well as the players $\{3,4\}$. The weight system $\mathbf{p}$ is asymmetric, and the the weighted pre-nucleolus as well as the weighted pre-kernel are given by $\nu^{\mathbf{p}}(N,v) = \{0,0,-1,1\}/2$ and do not satisfy {\bfseries ETP}. Let $\mathbf{y} = \nu^{\mathbf{p}}(N,v)$. Even though, we have $e^{v}(T,\mathbf{y}) = e^{v}(N\backslash T,\mathbf{y}) = 0$, and $y(T)=y(N\backslash T)=0$, we do not get that $\mathbf{y} = \mathbf{0}$ is drawn. Thus, from the solution $\nu^{\mathbf{p}}(N,v) = \{0,0,-1,1\}/2$, we realize that we cannot impose {\bfseries ETP} to conclude that from $y(T)=y(N\backslash T)=0$ the solution vector $\mathbf{y}$ must be the null-vector. Imposing {\bfseries ETP} as an assumption rather than a result from logical inference yields to a wrong conclusion. Furthermore, we derive $-3 = -p^{N}_{S} < s^{\mathbf{p}}_{kl}(\mathbf{y},v) = -3/2  = s^{\mathbf{p}}_{lk}(\mathbf{y},v) <  -p^{N}_{S^{\prime}} = -1$, contradicting what~\citeauthor{klep:13a} claim to show in their Theorem 3.3. Nevertheless, a further confirmation of their disproof.
\end{example}

\section{More misguided Logic from the Literature}  
\label{sec:2exp}
Unfortunately, the case discussed in the previous section is not the sole example of a mystified logic. A second case is the article of~\citet{watmut:08}. These authors try to study stable profit sharing in a patent licensing game while investigating licensing agreements in a bargaining set with a coalition structure. They employ, in almost all of their proofs by contradiction, the same line of logical wrong arguments as before. Similar, as in the preceding section, these authors also disprove their own results while being confused about propositional logic. As a consequence, at least Proposition 1, 2, 3, and 5 of~\citet{watmut:08} are invalid, and devalue their results. However, before we can go into the details, we have to introduce some additional notations and definitions from their article.

Let $N=\{1,\dots,n\}$ be a set of identical firms producing a homogeneous good. An external licensor called player $0$ has a patent  of a cost-reducing or quality improving technology. The set of players is $\{0\} \cup N$. Each non-empty subset of $\{0\} \cup N$ is a coalition. The game has three stages. At stage (i), the licensor selects a subset $S \subseteq N$ of firms to invite them in exclusive negotiation to acquire some licenses. In stage (ii) they negotiate about the payment made to the licensor. According to~\citeauthor{watmut:08}, this specifies at stage (iii) a TU game  with coalition structure denoted by $(\{0\}\cup N,v,P_{S})$, whereas $P_{S} = \{\{0\}\cup S\}\cup\{\{i\}\,\arrowvert\, i \in N\backslash S\}$. They assume, in addition, that whenever $s$ firms hold a license, then $W(s)$ denotes the competitive equilibrium gross profit of a licensee, and $L(s)$ the corresponding gross profit of a non-licensee. They require also that the following relations 
\begin{equation*}
  W(s) > L(0) \;\; \forall s=1,\ldots,n, \qquad L(0) > L(s) \;\; \forall s=1,\ldots,n-1,
\end{equation*}
hold. From this, a characteristic function $v:2^{\{0\}\cup N} \mapsto \mathbb{R} $ is defined through
\begin{equation*}
  v(\{0\}) = v(\emptyset) = 0, \quad v(\{0\}\cup T) = t\,W(T),\quad v(T) = t\,L(\rho(t)), \quad \forall \emptyset \neq T \subseteq N,
\end{equation*}
whereas $L(\rho(s)) := \min_{r=\arrowvert R \arrowvert, R \subseteq N\backslash S}\,L(r)$.

\citeauthor{watmut:08} define the set of imputations for all permissible coalition structures $P_{S}$ as
\begin{equation*}
 \begin{split}
  X_{S} := \big\{ \mathbf{x} & = \{x_{0},x_{1},\ldots,x_{n}\} \in \mathbb{R}^{n+1}\,\big\arrowvert\,x_{0} + \sum_{i \in S}\,x_{i} = s\, W(s), \\ 
   & x_{0} \ge 0, x_{i} \ge L(\rho(1))\;\forall i \in S,\; x_{j} = L(s)\;\forall j \in N\backslash S \big\}.
 \end{split}
\end{equation*}

They define, in addition, the core of a game with a coalition structure $P_{S}$ as a subset of $X_{S}$ which is given by
\begin{equation*}
  C_{S}=\big\{\mathbf{x} \in X_{S}\,\arrowvert\, x(T) \ge v(T) \quad\forall\,T \subseteq \{0\} \cup N, \quad T \cap (\{0\} \cup S) \neq \emptyset \big\}.
\end{equation*}

The bargaining set w.r.t. a coalition structure $P_{S}$ is defined by
\begin{equation*}
  M_{S}=\big\{\mathbf{x} \in X_{S} \;\arrowvert\; \text{no player in } \{0\} \cup S \;\text{has a valid objection at }\,\mathbf{x}\,\big\}.
\end{equation*}

Then the following symmetric solutions are defined by 
\begin{equation*}
  \tilde{C}_{S} = C_{S} \cap \tilde{X}_{S}, \qquad \tilde{M}_{S}= M_{S} \cap \tilde{X}_{S},
\end{equation*}
where $\tilde{X}_{S}=\{\mathbf{x} \in X_{S}\,\arrowvert\, x_{i} = x_{j} = \tilde{x}\,\forall i,j \in S\}$.

 The argumentation of~\citeauthor{watmut:08} is best observed by Proposition 1. There, those authors argue by an indirect argument that if $A \land \neg B$ is valid, then $B$ follows, hence, a contradiction is drawn to infer that $A \Rightarrow B$ must be given. Similar as above, these authors conclude from a wrong implication $A \land \neg B \Rightarrow B \land \neg B$ that the logical equivalent statement $A \Rightarrow B$ is satisfied. Nevertheless, both statements are false, disproving their Proposition 1. Once more, the crucial arguments are set in italic and highlighted in red.
\begin{quote}
  \begin{labeling}[:]{Proposition}
  \item[\bfseries{Proposition 1~(\citet[p. 512]{watmut:08})}]  $C_{S} = \emptyset$ if $S \neq N$.
  \end{labeling}
  \begin{proof}
    We first show that $\tilde{C}_{S} = \emptyset$ if $S \neq N$. {\itshape \color{red}  Suppose $\tilde{C}_{S} \neq \emptyset$. Take $\mathbf{x} \in \tilde{C}_{S}$} with $x_{i} = \tilde{x}$ for any $i \in S$. If $\tilde{x} \le L(0), \sum_{i \in N}\,x_{i} = s\,\tilde{x}+(n-s)\,L(s) < n\,L(0) = v(N)$ because $L(0) > L(s)=x_{j}$ for any $j \in N\backslash S$. Hence, $\tilde{x} > L(0)$. {\itshape \color{red} Next take a coalition $\{0\} \cup T$ such that $\arrowvert T \arrowvert = \arrowvert S \arrowvert$, $T \subseteq N \backslash S$ if $\arrowvert S \arrowvert \le n/2$ and $T \supseteq N\backslash S$ if $\arrowvert S \arrowvert > n/2$}. Let $t=\arrowvert T \arrowvert$. Then $x_{0}+\sum_{i \in T}\,x_{i} < s\,W(s) = t\,W(t)$, because $x_{0}+s\,\tilde{x} = s\,W(s)$ and $\tilde{x}>L(0)>L(s)$. {\itshape \color{red} This contradicts $\mathbf{x} \in \tilde{C}_{S}$. Finally, $\tilde{C}_{S} = \emptyset$ implies $C_{S} = \emptyset$} by Lemma 1.~(\citet[p. 512]{watmut:08})
  \end{proof}
\end{quote}
\citeauthor{watmut:08} try to show that 

if $S \neq N$ (A is true), then $C_{S} = \emptyset$ (B is true). 

\noindent The contrapositive of this statement is given by

if $C_{S} \neq \emptyset$ (B is false), then $S = N$ (A is false).

\noindent They start by assuming that $\tilde{C}_{S} \neq \emptyset$, hence $C_{S} \neq \emptyset $ ($\neg B$) is satisfied. In the next step, they construct a vector from $\tilde{C}_{S}$. For doing so, they attain that $S \neq N$ is given ($A$ is true) due to the construction of $S$, to finally conclude that $\mathbf{x} \notin \tilde{C}_{S}$, from which $C_{S} = \emptyset$ ($B$) is attained by those authors. The authors want to employ an indirect proof while drawing from a valid assumption to a contraction. But, we observe again that they have actually shown that from a truth $A \land \neg B$ one derives $B$. But this means that they deduce from the wrong implication $A \land \neg B \Rightarrow B \land \neg B$ that the implication {\itshape ``if $S \neq N$, then $C_{S} = \emptyset$''} ($A \Rightarrow B$) is given, this is a fallacy due to $(A \land \neg B \Rightarrow B \land \neg B) \equiv (A \Rightarrow B)$. They have incorrectly applied $(\phi \Rightarrow \bot) \Leftrightarrow \neg \phi$. In fact, they have established that $A \Rightarrow B$ is an invalid implication, disproving their own proposition. Similar as by~\citeauthor{klep:13a}, they also being confused by propositional statements.

By investigating the proof of Lemma 2 from~\citet[p. 514]{watmut:08}, we also have to realize that this kind argumentation was not an isolated event. Glancing through the whole article, we observe that those authors have applied this fallacy several times, since almost all of their results are false. By studying their arguments for proving the Lemma 2, we find the same wrong usage of the indirect proof as for their proof of Proposition 1. Instead of assuming that $A \land \neg B$ is invalid to deduce that a contradiction follows in order to get that $(\neg A \lor B) \equiv (A \Rightarrow B)$ is valid, they argue that a truth $A \land \neg B$ implies a falsehood $\neg A$, from which they infer that $A \Rightarrow B$ follows.

\begin{quote}
  \begin{labeling}[:]{Lemma}
  \item[\bfseries{Lemma 2~(\citet[p. 514]{watmut:08})}]  For any $S \subseteq N$, if $\mathbf{x} \in \tilde{M}_{S}$ then $x_{0} \le s^{*}\,(W(s^{*}) - L(0))$.
  \end{labeling}
  \begin{proof}
    Let {\itshape \color{red} $\mathbf{x} \in \tilde{M}_{S}$}. Suppose {\itshape \color{red} $x_{0} > s^{*}\,(W(s^{*}) - L(0))$}. By the definition of $s^{*}, \tilde{x}=(s\,W(s)-x_{0})/s < (s\,W(s)-s^{*}\,(W(s^{*}) - L(0)))/s \le L(0)$. Take an objection $(\mathbf{y},N)$ of $i \in S$ against the licensor in $\mathbf{x}$ with $y_{k}=L(0)$ for any $k \in N$. If the licensor had a counter objection $(\mathbf{z},\{0\}\cup T)$ to the objection with $z_{0} \ge x_{0} > s^{*}\,(W(s^{*}) - L(0))$ and $z_{k} \ge y_{k}=L(0)$ for any $k \in T$, it should be $z_{0} + \sum_{k \in T}\,z_{k} > s^{*}\,(W(s^{*}) - L(0)) + t\,L(0) \ge t\,W(t)$ by the definition of $s^{*}$, where $t=\arrowvert T \arrowvert $. Hence, no counter objection can be made, {\itshape \color{red} contradicting that $\mathbf{x} \in \tilde{M}_{s}$}.~(\citet[p. 514]{watmut:08})
  \end{proof}
\end{quote}
Again, \citeauthor{watmut:08} try to apply an indirect proof based on a material implication to the statement

if $\mathbf{x} \in \tilde{M}_{S}$ ($A$ is true), then $x_{0} \le s^{*}\,(W(s^{*}) - L(0))$ ($B$ is true),

\noindent which is equivalent to the contrapositive 

if $x_{0} > s^{*}\,(W(s^{*}) - L(0))$ ($\neg B$), then $\mathbf{x} \not\in \tilde{M}_{S}$ ($\neg A$).

\noindent In their proof, the authors have slightly changed their line of argument while supposing first that $\mathbf{x} \in \tilde{M}_{S}$ ($A$ is true), and by the next step that $x_{0} > s^{*}\,(W(s^{*}) - L(0))$ ($\neg B$) is satisfied, to finally conclude that $\mathbf{x} \not\in \tilde{M}_{S}$ ($\neg A$) must follow. By the same reasoning as above, this argumentation is logically false, since they have shown the wrong implication $A \land \neg B \Rightarrow A \land \neg A$. Again,~\citeauthor{watmut:08} have disproved their own Lemma 2, as a consequence, the statement {\itshape ``if $\mathbf{x} \in \tilde{M}_{S}$ ($A$), then $x_{0} \le s^{*}\,(W(s^{*}) - L(0))$ ($B$)''} does not hold. 

Proposition 3 of~\citet{watmut:08} is false, since Lemmata 2, 4, and 5 are not correct, and therefore Proposition 5 is false either. The reader will observe while inspecting these purported proofs in more detail that those authors have again disproved themselves with the consequence that this devalues the whole article.

We close this section while mentioning a third case where an author deduces wrong conclusions from logical statements derived from an indirect proof which relies on a material implication. We only summarize the main arguments by the author without going into the details, and without discussing the notation as well as the definitions.

In the article of~\citet{lard:12}, the author claims to provide for the class of oligopoly TU games an existence result of the $\gamma$-core and a single-valued allocation rule inside of the $\gamma$-core that is called by the author Nash Pro rata-value. Moreover, ~\citet{lard:12} asserts to present an axiomatic characterization of the NP-value. However, even this article is false due to the fact that the author confuses and mixes up non-equivalent fundamental statements from propositional logic in applying false indirect arguments.~\citeauthor{lard:12} neither recognizes the logical relationship $(A \land \neg B \Rightarrow A \land \neg A) \equiv (A \Rightarrow B)$ nor $(\neg A \land B \Rightarrow \neg A \land A) \equiv (B \Rightarrow A)$.  

His proof of the {\itshape ``sufficiency case''} of Proposition 3.1 is not correct. Similar as in the other examples, he uses elements from a material implication for establishing the logical equivalent proposition {\itshape if $A$ then $B$}. This author starts with $A \land \neg B$ to perform this kind of proof to get a contradiction in order to conclude that the implication $A \Rightarrow B $ is drawn. Once more, this author does not recognize that whenever a valid premise $A \land \neg B$ implies something false like $\neg A$, one cannot get a true statement. In this case, the implication must be falsehood. Similar to the other cases, this author applies the prerequisite $A$ of the positive statement and $\neg B$ in order to prove the contrapositive statement {\itshape if $\neg B \Rightarrow \neg A$}. First, he assumes that the payoff vector $\hat{x}^{\mathcal{P}} \in X^{\mathcal{P}}$ is a Nash equilibrium of the normal form oligopoly game $\Gamma^{\mathcal{P}} = (\mathcal{P}, (X^{S},\pi_{S})_{S \in \mathcal{P}})$, that is, premise $A$ holds, and then assuming in the next step that the strategy profile $\hat{x} = (\hat{x}_{S})_{S \in \mathcal{P}} \in X_{N}$ is not a Nash equilibrium of the normal form oligopoly game $\Gamma = (N, (X_{i},\pi_{i})_{i \in N})$ under $\mathcal{P}$, i.e., premise $B$ is false. Premise $A$ is then used in his proof to construct in a first step the vector $\hat{x}$, and finally to construct the contradiction that $\hat{x}^{\mathcal{P}} \in X^{\mathcal{P}}$ is not a Nash equilibrium ($\neg A$). In effect, he has shown that $A \land \neg B \Rightarrow A \land \neg A$ is a wrong proposition. As a consequence, the implication $A \Rightarrow B$ must be false too, in accordance with $(A \land \neg B \Rightarrow A \land \neg A) \equiv (A \Rightarrow B)$. The author incorrectly applied $(\phi \Rightarrow \bot) \Leftrightarrow \neg \phi$.

For completeness, we just want to mention that the same misguided line of argument is also given for the {\itshape ``necessity case''}. There, he is not aware about the following logical equivalence $(\neg A \land B \Rightarrow \neg A \land A) \equiv (B \Rightarrow A)$. No wonder that he shows that the truth $\neg A \land B$ implies a falsehood $\neg A \land A$, which is as well a wrong implication. It follows that $B \Rightarrow A$ must be invalid. In summary, he has shown in both cases the exact opposite of what he had claimed to prove. As a consequence,~\citeauthor{lard:12} has disproved his own Proposition 3.1.    
 
In the sequel, we show what will happen if we apply a proof by contraposition $\neg B \Rightarrow \neg A$ for the {\itshape ``sufficiency case''} in order to see where we run into problems. But then the starting point of the proof has to be the assumption that the payoff vector $\hat{x} \in X_{N}$ is not a Nash equilibrium of the normal form oligopoly game $\Gamma = (N, (X_{i},\pi_{i})_{i \in N})$ under $\mathcal{P}$ ($\neg B$), which implies by imposing the correct assumption like quasi-concavity on the profit function $\pi_{i}$ in order to guarantee existence of an equilibrium that 
\begin{equation*}
  \sum_{i \in S}\, \pi_{i}(\hat{x}_{S},\hat{x}_{-S}) \le   \sum_{i \in S}\, \pi_{i}(\check{x}_{S},\hat{x}_{-S}),
\end{equation*}
is true. In this case, Formula (11) of~\citet[p. 394]{lard:12} implies for payoff vector $\hat{x} \in X_{N}$ that only
\begin{equation*}
  \sum_{i \in S}\, C_{i}(\hat{x}_{i}) \ge C_{S}(\hat{x}^{S}).
\end{equation*}
can be estimated, since it cannot be supposed that $\hat{x}^{\mathcal{P}} \in X^{\mathcal{P}}$ is a Nash equilibrium. As a consequence, it is also not anymore clear that 
\begin{equation*}
  \pi_{S}(\hat{x}^{\mathcal{P}}) < \pi_{S}(\check{x}^{S}, \hat{x}^{-S}),
\end{equation*}
is satisfied as it was claimed by~\citet[p. 395]{lard:12}. This inequality can only be obtained when the author can establish by some logical inference that $\hat{x}^{\mathcal{P}} \in X^{\mathcal{P}}$ is a Nash equilibrium of the normal form oligopoly game $\Gamma^{\mathcal{P}} = (\mathcal{P}, (X^{S},\pi_{S})_{S \in \mathcal{P}})$ ($A$ is valid), but not by an assumption. Moreover, Corollary 3.2 is not correct either, implying in connection with the disproof of Proposition 3.1 that the TU game in $\gamma$-characteristic function form is not well-defined. Again, the results of the article are devalued according to these logical flaws.   

\section{Concluding Remarks}
\label{sec:rem}
We have demonstrated on a small sample from the game theory literature, how fatal it can be for the reliability of the derived results, when authors have not imposed a simple and quick logical cross-check on their argumentation. We focused on the indirect proof based on a material implication to report some logical failures committed in the literature, and how we have to proceed in order to get logical correct propositions. Even though ostensible, the derived results seem to be sound and rigorous, they are, nevertheless, wrong, since they have violated fundamental statements from propositional logic. In fact, we observed that these authors have disproved themselves, invalidating the results and articles.

\pagestyle{scrheadings} \chead{\empty}  
%%\small
\footnotesize
%\pdfbookmark[1]{References}{bib}
\bibliography{indproof}%\label{bib}

\end{document}